| Title | Reactive oxygen species trigger downward vertical migration in diatom microphytobenthic biofilms as a strategy to cope with oxidative stress |
|---|---|
| Authors | Alexandre Desparmet[1], Bruno Jesus[2], Tony Robinet[1], Thierry Dufour[3], Cédric Hubas[1] |
| Affiliations | [1] Laboratoire de Biologie des Organismes et Écosystèmes Aquatiques (BOREA), UMR8067, Muséum National d'Histoire Naturelle, Sorbonne Université, CNRS, IRD, UCN, UA, Station Marine de Concarneau, Quai de la Croix, Concarneau 29900, France<br>[2] Institut des Substances et Organismes de la Mer (ISOMer), UR2160, Nantes Université, Faculté des Sciences et Techniques, Nantes 44000, France<br>[3] Laboratoire de Physique des Plasmas, 4 Place Jussieu, 75005 Paris, Sorbonne Université, CNRS, Ecole polytechnique. |
| Correspondence | alexandre.desparmet@mnhn.fr, thierry.dufour@sorbonne-universite.fr, cedric.hubas@mnhn.fr |
| Ref. | The ISME Journal, 2026, 20(1), wrag034 |
| DOI | https://doi.org/10.1093/ismejo/wrag034 |
| Summary | Diatom-dominated intertidal microphytobenthic biofilms experience daily fluctuations in irradiance, which can lead to oxidative stress within the photosynthetic apparatus through the production and accumulation of reactive oxygen species. To maintain photosynthetic efficiency, benthic diatoms have developed protective strategies, including mobilization of the antioxidant xanthophyll cycle and the ability to migrate vertically through sediments. However, mechanistic understanding of signaling pathways underlying migration remains poorly characterized. This study investigated the triggering effect of reactive oxygen species on behavioral and photophysiological responses through the analysis of lipophilic pigments and fluorescence parameters. To this end, two microphytobenthic communities, one with sediment allowing vertical migration and another without sediment restricting it, were exposed to irradiance, cold atmospheric plasma, and hydrogen peroxide stresses. Results showed a consistent downward migration response under all oxidative stresses, highlighting the key role of reactive oxygen species, especially hydrogen peroxide, in triggering this microphytobenthic behavior. Moreover, a difference was observed between the pathways involved in vertical migration and those underlying photoprotective responses. Hydrogen peroxide and cold atmospheric plasma stresses highlighted the necessity for substantial microphytobenthic migration, whereas irradiance induced a specific and controlled response involving engagement of the xanthophyll cycle, acting in synergy with the migration strategy by showing stronger activation when migration was impaired. By establishing that a rapid and efficient migration could be induced by reactive oxygen species (ROS) and could act in synergy with the xanthophyll cycle in epipelic cells, this study provides key insights into the molecular basis of microphytobenthic responses to cellular and environmental oxidative stresses. |
| Keywords | diatom microphytobenthic biofilms, Pleurosigma strigosum, vertical migration, ROS, light stress, cold atmospheric plasma, hydrogen peroxide, xanthophyll cycle, photophysiology |

# 1. Introduction

The aquatic microbiome exhibits a remarkable diversity of adaptive strategies to deal with rapid environmental fluctuations. Intertidal sediments, as dynamic interfaces, undergo significant qualitative and quantitative variations in the electromagnetic spectrum throughout the day, impacting the phototrophic communities [1]. Among these, microphytobenthos dominated by epipelic diatoms, forms structured mucilaginous biofilms on muddy substrata, an advantageous lifestyle that plays a key role in their adaptation and associated ecosystem services [2-4].

Light carries essential environmental information that diatoms can perceive either directly through photoreceptors or indirectly via cellular signals, such as levels of reactive oxygen species (ROS) and the redox state of the photosynthetic chain [5-7]. These signals trigger physiological responses involved in light acclimation, photoprotection and photomotile movements, enabling diatoms to acclimate, adapt, and survive [8]. Under high light intensity conditions, photosynthetic capacity can be exceeded, leading to the synthesis and accumulation of ROS, natural by-products of oxygen metabolism, including singlet oxygen $^1O_2$, superoxide $O_2^{\bullet-}$, hydrogen peroxide $H_2O_2$ (e.g. via the Mehler reaction), and hydroxyl $HO^\bullet$ [9]. This deregulation disrupts homeostasis and causes photooxidative damage, such as lipid peroxidation, which compromises cellular integrity [10], and the oxidation of DNA and key proteins such as the photosystem II D1 protein [11,12], which can ultimately result in cell death [9,13]. Epipelic diatoms have developed two major adaptive strategies to preempt this ROS formation and accumulation, ensuring optimal photosynthetic processes in a variable light environment: vertical migration behavior and xanthophyll cycle [14,15].

First described in 1907 by Fauvel and Bohn, the ability of cells to move vertically through muddy substrata is a behavior regulated both endogenously and plastically by various environmental forcing, including light information [16-21]. Rapid and temporally adjustable, this actin-myosin-driven gliding system involving the secretion of extracellular mucilage fibrils through the slit raphe, enables diatoms to quickly reposition themselves at low metabolic cost, allowing them to reach sediment layers with optimal irradiance conditions [20-26]. This adaptation, partly responsible for the diversity of Bacillariophyta [27], represents a significant evolutionary advantage and has given rise to distinct migration strategies depending on species, previous light photoacclimation, and the environmental context [15,17,19,20]. In addition, the optimization of diatom light harvesting results from the dynamic regulation of accessory carotenoid pigments, such as xanthophylls linked with the photosystem II Fucoxanthin-Chlorophyll-Protein supercore complex [6,14,28,29]. Overstimulation of the photosynthetic electron transport chain acidifies the thylakoid lumen through proton translocation associated with the cytochrome b6f complex via the Q-cycle, as well as proton release generated by water photolysis at photosystem II. Under these low-lumen-pH conditions, the enzymatic activity of Diadinoxanthin de-epoxidase is upregulated, leading to the de-epoxidation of Diadinoxanthin embedded in the thylakoid membranes [15,29-31]. Rapid xanthophyll adjustments involved in the energy-dependent component of non-photochemical quenching (qE) are particularly pronounced in diatoms. This represents a crucial antioxidant photoprotective response that optimizes thermal dispersion of excess energy in a fluctuating environment, thereby limiting ROS-associated damage in the chloroplast [15,29-31]. These two major adaptations are complemented by others, such as chloroplast mobility [32,33], spatial cell orientation [34,35], frustule nanostructures [36], and cellular ROS management through enzymatic and non-enzymatic antioxidant capacities [31,37-39]. Together, these adaptations form a sophisticated system that enables diatoms to monitor and respond to environmental fluctuations, contributing to their success in modern marine ecosystems [6].

Although diatom photoregulation via vertical migration and the xanthophyll cycle are both well characterized individually, the mechanistic understanding of molecular ROS-related cellular signaling pathways that coordinate and integrate real-time protective strategies remains poorly understood in benthic diatoms. The present study, conducted on a natural epipelic microphytobenthic biofilm, addresses this gap by testing the hypothesis that: (i) ROS can act as a trigger signal for downward vertical migration, and (ii) ROS-induced migration may be initiated through pathways that operate independently of photoprotection.

# 2. Materials & Methods

## 2.1. Biofilm sampling and experimental treatments

### Microphytobenthos sampling

Microphytobenthos was collected from a disused outdoor breeding pond at the Concarneau Marine Station (France; 47°52’05.85”N; 3°55’00.51”W). The pond retains at least 5 cm of seawater above the sediment at low tide, ensuring that the biofilm developing on fine-grained muddy sediments under a semi-diurnal tidal regime remains permanently immersed (**Fig. 1A** and **Fig. 1B**). The upper centimeter of sediment was sampled at low tide on 29 April 2024 (1:00 p.m.) using a shovel. Muddy samples were placed in plastic boxes with overlying seawater and left to settle under natural ambient laboratory light conditions, without direct illumination, for 24 h at 20°C.

### Community and replicate conditioning

Two communities were compared: one capable of vertical migration through sediment, exhibiting a robust and regular pattern driven by both endogenous circadian and tidal rhythms (sediment-associated biofilm), and another in which vertical movement was impaired due to the absence of sediment (sediment-free biofilm). Petri dishes (5.5 cm diameter) containing sediment-associated biofilms were prepared 24 h before the experiment, whereas Petri dishes containing sediment-free biofilms were prepared on the day of the experiment from biofilm that had settled in a box for 24 h. The sediment-associated biofilms consisted of homogeneous sediment discs (5 mm thick) topped with 4 mL of filtered seawater from Concarneau Bay, which were left to settle for 24 h under ambient laboratory light and temperature. The sediment-free biofilms were prepared by gently peeling off the highly cohesive, EPS-rich upper millimeter of the microphytobenthos with tweezers and resuspending it in filtered site seawater (ratio 1 mg: 20 mL). This method minimizes mechanical disturbance and preserves the surface community’s cell-size distribution and taxonomic composition. Four milliliters of the sediment-free biofilm suspension were placed in each Petri dish.

### Oxidative treatments

Both sediment-associated and sediment-free biofilm communities were exposed in the morning (9:00 a.m. to 1:00 p.m.), during the biofilm’s upward migration phase, to three short oxidative stress treatments. These were compared with dark-adapted controls (<5 µmol.photons.$m^{-2}.s^{-1}$, which is not complete darkness, but remains below the light compensation point for microphytobenthos [40]). The high-light stress was applied to mimic intense irradiance commonly experienced by autotrophic cells in natural environments, with a 30-min exposure to 1500 µmol.photons.$m^{-2}.s^{-1}$ of full-spectrum warm white LED light panel (**Supplementary Fig. 1A**). A cold atmospheric plasma was applied for 10 minutes by piezoelectric discharge, using ambient laboratory air as the carrier gas. Exposure was performed indirectly using plasma-activated seawater, with a 3 mm gap maintained between the water surface and the plasma source to prevent mechanical disturbance of diatoms from transient thermal filaments. This treatment generates both reactive oxygen and nitrogen species (e.g. $H_2O_2$, $NO_2^-$, and $NO_3^-$), inducing a composite oxidative stress predominantly driven by $H_2O_2$, which reaches a concentration of 209 µM (additional plasma details provided in the Supplementary materials and methods). Finally, we applied a targeted hydrogen peroxide stress, consisting of a 10-min exposure to a 200 µM $H_2O_2$ seawater solution, allowing the specific effect of this ROS to be isolated. The $H_2O_2$ concentration was determined based on the amount produced during the 10-min plasma treatment, ensuring that the hydrogen peroxide and plasma stresses were as comparable as possible. Both hydrogen peroxide and plasma exposures were conducted in darkness. For sequencing-based community characterization and pigment analyses, five biological replicates (n=5) per community and treatment, all originating from the same initial sampling, were immediately frozen with liquid nitrogen after exposure, lyophilized, and stored at −80°C. Three additional replicates (n=3) were dedicated to photophysiological parameter measurements (**Table 1**). Additional details on biofilm sampling and experimental treatments are provided in the Supplementary materials and methods.

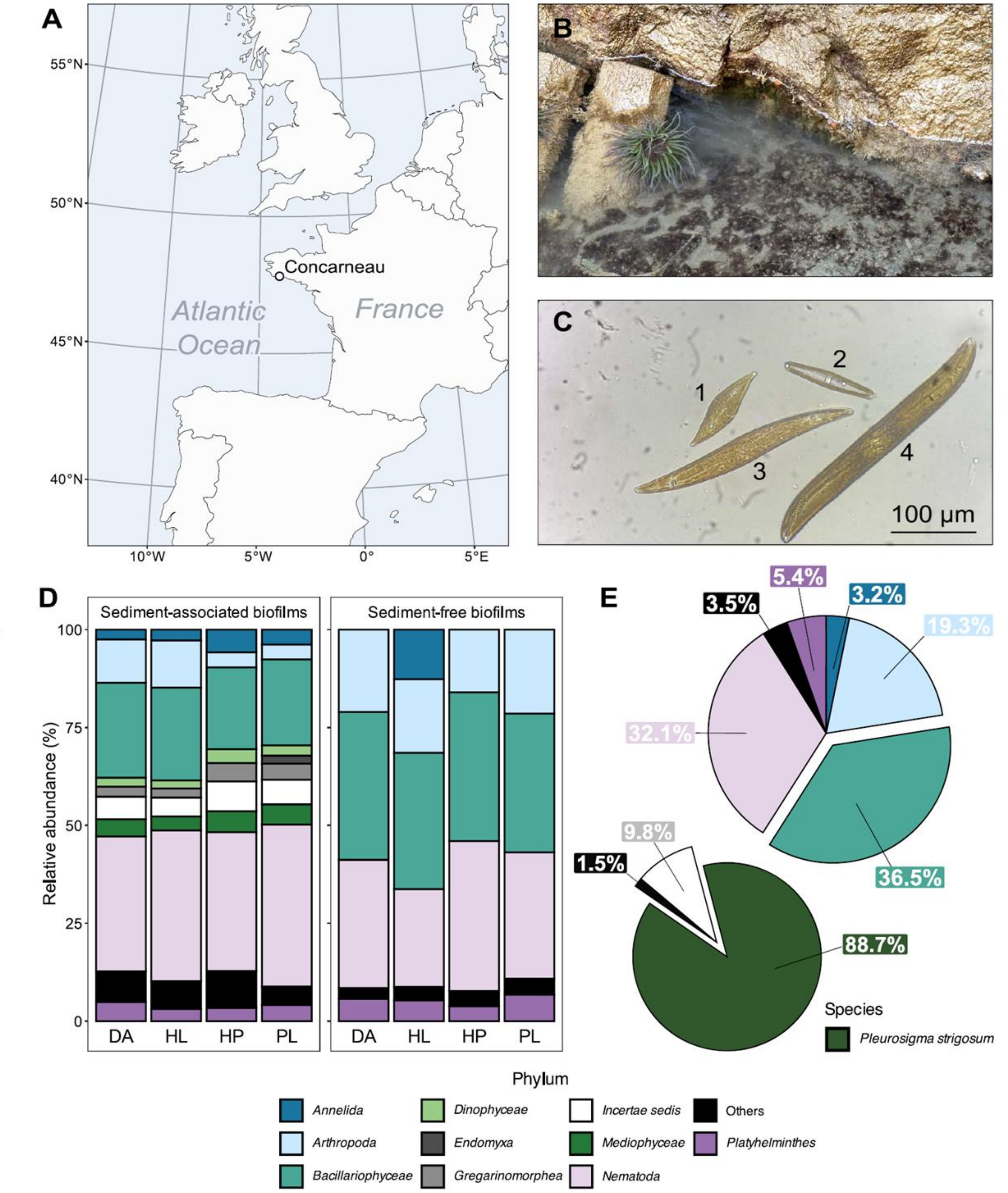

***Figure 1 MICROPHYTOBENTHIC COMMUNITY DIVERSITY. [A] Geographical location of the studied microphytobenthic biofilm. [B] Microphytobenthic biofilm (brown patches) developing on muddy substrata of the breeding pond. [C] Light microscopy image of the most commonly observed diatom species in the microphytobenthic community: 1. Pleurosigma sp., 2. Navicula sp., 3. Pleurosigma strigosum, 4. Gyrosigma sp.. [D] Relative average abundance barplots (%; n=5 per treatment) of eukaryotic phyla from microphytobenthic samples (some sequences derive from planktonic eDNA, thus contributing to the observed community composition) for sediment-associated and sediment-free biofilms, under dark-adapted control (DA), high-light (HL), hydrogen peroxide (HP), and cold atmospheric plasma (PL) treatments; [E] Average sediment-free microphytobenthic biofilm community across all treatments, highlighting the prevalence of the autotrophic phylum Bacillariophyceae (first pie chart), and the significant contribution of P. strigosum within Bacillariophyceae abundances (second pie chart). The "others" category includes the sum of taxa with <2% abundance.***

*Table 1. Experimental setup. Summary of the materials and treatments tested in this study.*

| Treatments | Dark-adapted control | High-light | Cold atmospheric plasma | Hydrogen peroxide |
|---|---|---|---|---|
| Description | Negative control | Natural behavior control | Experimental composite treatment | Positive control |
| Stress type | - | Warm white UV to near-IR light | Reactive oxygen and nitrogen species stress | Targeted oxidative stress ($H_2O_2$) |
| Time exposure | - | 30 min | 10 min | 10 min |
| ROS origin | - | Endogenous | Exogenous | Exogenous |
| Experimental materials | Leave in darkness | LED panel Rhino Series (GR-RHINO260-PRO-DIM) | Plasma activated seawater (PiezoBrush PZ3 with NearField insert from Reylon Plasma GmbH), leave in darkness | Seawater concentrated with $H_2O_2$ (ref. 216763 from Sigma Aldrich), leave in darkness |
| Molecular analyses | Sediment-associated biofilm, n=5; Sediment-free biofilm, n=5 | | | |
| Photophysiological measurements | Sediment-associated biofilm, n=3; Sediment-free biofilm, n=3 | | | |

## 2.2. 18S rRNA gene amplicon sequencing

To identify the motile diatoms involved in vertical migration behavior, the eukaryotic microphytobenthic community was characterized through a short-read sequencing on a NovaSeq 6000 System (Illumina). DNA was extracted using the NucleoSpin Soil Mini kit (MACHEREY-NAGEL, ref. 740780.10), and the V4 region of the 18S rRNA gene was amplified using primers TAReuk454FWD1 and TAReukREV3 **[41]**. Library construction and sequencing were performed by the company Biomarker Technologies (BMKGENE) GmbH (Münster, Germany).

## 2.3. Chlorophyll fluorescence: pulse amplitude modulated fluorometry

In order to monitor cellular efficiency under the different treatments and interpret subsequent photoprotective responses, photosynthetic parameters derived from Chlorophyll a fluorescence were measured using a Monitoring Pen MP 100-E (Photon System Instruments). The light curve program consisted of seven 1-min irradiance steps, increasing from 10 to 1000 µmol.photons.m$^{-2}$.s$^{-1}$ ($\lambda_{LED}$ excitation = 470 nm), with consistent gain, superpulse, and flash pulse intensity settings maintained across all measurements. The relative electron transport rate (rETR) of photosystem II, an indicator of productivity **[23]** and in some cases a proxy for carbon fixation **[42]**, corresponds to the values extracted directly from light curve points. The initial linear α-slope, representing the maximum light-use efficiency of photosystem II, was derived from rETR-irradiance (rETR-E) curves using a previously described and modified model **[43,44]**. The maximum quantum yield of photosystem II photochemistry under dark-adapted state ($F_v/F_m$), as well as non-photochemical quenching (NPQ) and its regulated Y(NPQ) and non-regulated Y(NO) components, were calculated using established equations [45,46] (**Supplementary Table 1**). All fluorescence measurements of photosynthetic parameters for each sample and treatment were performed twice: once before and once after stress exposure, each time following a minimum dark adaptation period of 10 min. After the high-light exposure, post-stress measurements were taken after this dark adaptation period once the exposure had ended, whereas for the hydrogen peroxide and plasma, measurements were performed immediately after the 10-min stress period, as the samples were already dark-adapted under these treatments. For the dark-adapted control samples, which do not receive any stress, measurements were performed in parallel with the experimental samples, ensuring that they were in the same temporal framework.

## 2.4. Monitoring of vertical migration

To assess the vertical migratory response over time, the minimum fluorescence of dark-adapted samples ($F_0$), used as a proxy of surface algal biomass **[14,17]**, was monitored in the sediment-associated biofilm community after a 24-h settling period under ambient laboratory light and temperature. $F_0$ measurements were conducted independently and randomly on dense biofilms (182 cm$^2$) contained in 14 × 13 cm boxes (one box per stress), then compared with their respective dark-adapted biofilms (high-light and plasma: n=8; hydrogen peroxide: n=5). Due to the rapid induction of the migratory response in microphytobenthos, $F_0$ biomass monitoring was resumed after 5 min of dark adaptation following the end of the high-light stress, and immediately after the hydrogen peroxide and plasma stresses.

## 2.5. Lipophilic pigments analysis

To characterize the metabolic impact of the treatments and complement the photosynthetic efficiency data, the lipophilic pigments were extracted according to a previously published method **[47]**. Freeze-dried samples (10 mg of sediment-free biofilm and 40 mg of sediment-associated biofilm) were incubated for 15 min at −21°C in 2 mL of 95% methanol buffered with 2% ammonium acetate, then filtered through a 0.2 µm PTFE filter. Aliquots of 100 µL were injected into an Agilent 1260 Infinity HPLC system equipped with a C18 Supelcosil column and a UV-VIS photodiode array detector (DAD 1260 VL, 250-900 nm). The method employed a flow rate of 0.6 mL.min$^{-1}$ using solvent mixture A (0.5 M ammonium acetate in 85:15 methanol: water), B (90:10 acetonitrile: water), and C (100% ethyl acetate). A total of 44 microphytobenthic pigments were detected (list and abbreviations are provided **Supplementary Table 2**), and their concentrations (µg pigment.g$^{-1}$ dry weight) were quantified using calibration curves based on phytoplankton pigment standards from DHI. Two groups of derived Chlorophyll a were identified: the “Chlorophyll a-derivatives” group includes unidentified Chlorophyll a-like pigments, as well as the allomer, and epimer forms; the “Pheopigments” group includes Pheophytins a, Pheophorbides a, and Pyropheophytin a. The contribution (%) of each Chlorophyll a-derivatives and Pheopigments group to the total Chlorophyll a pool was calculated by dividing their concentrations by the $\sum[Chl\ a]$ concentration, used as a proxy of biomass:

$$\sum[Chl\ a] = [Chl\ a] + \begin{bmatrix} Chl\ a \\ derivatives \end{bmatrix} + [Pheopigments]$$

The de-epoxidation state (%), defined as the xanthophyll ratio of Diatoxanthin to the Diadinoxanthin + Diatoxanthin pool, was calculated as follows:

$$\begin{pmatrix} Deepoxidation \\ state_{\%} \end{pmatrix} = 100.\frac{[Diatoxanthin]}{[Diadinoxanthin] + [Diatoxanthin]}$$

Additionally, pigment concentrations were used to reconstruct the microphytobenthic community absorption spectrum for Qphar estimation, thus providing a more accurate assessment of light actually absorbed by diatoms at the substrata surface during the high-light treatment (**Supplementary Figs 1B-E**; see **Supplementary materials and methods**).

## 2.6. Data treatment

Raw 18S rRNA gene amplicon sequencing data were processed via the SAMBA workflow to generate amplicon sequence variants, which were assigned taxonomically with the PR2 (v5.0.0) database. Reads were normalized by coverage-based rarefaction and transformed to relative abundances. Diversity between communities was tested using the Wilcoxon rank-sum test.

For both communities, the net impact values of stress ($\Delta_{\text{After-Before}}$) on photophysiological parameters were calculated and compared to those of dark-adapted control samples, which accounts for metabolic drift over time, using Welch’s two-sample t-test.

Pigment composition was analyzed using ANOSIM tests based on Euclidean distances between individual coordinates. These were extracted from the first two axes of each PCA performed on Hellinger-transformed pigment concentrations to assess the significant effects of treatments and communities. A Between-Class Analysis was performed to characterize the effects of the stress factor on the total inertia of pigment profiles for both communities. A SIMPER analysis on relative proportions, performed independently for each sediment-associated and sediment-free biofilm community, was used to identify pigments that varied significantly under each stress compared to the dark-adapted control samples. Changes in pigment ratios were tested using the Mann-Whitney U test.

# 3. Results

## 3.1. Motile diatom *Pleurosigma strigosum* dominates the eukaryotic microphyto-benthic community

Despite inherent biases, such as the amplification of free eDNA or the underrepresentation of rare species in communities **[48]**, the sequencing approach provided a relevant semi-quantitative overview of the biovolume-dominant eukaryotic microphytobenthic taxa. The two communities displayed distinct assemblages, with significantly lower diversity indices and reduced variability among sediment-free biofilm replicates compared with sediment-associated biofilms (Wilcoxon-Mann-Whitney, $P<.05$) (**Supplementary Fig. 2**). The eukaryotic community was dominated by Bacillariophyceae, Nematoda, and Arthropoda, with an expected increase in the relative proportion of Bacillariophyceae in sediment-free biofilms, due to the substrata surface layer subsampling protocol which excluded much of the sediment-associated diversity (**Fig. 1D**). Autotrophic biovolume was largely dominated by the raphid species *P. strigosum*, which accounted for 88.7% of sediment-free biofilm relative abundances, together with other benthic diatoms, such as the genera *Navicula* and *Gyrosigma* (possibly *Gyrosigma balticum* **[49]**) (**Fig. 1C, Fig.1E**).

## 3.2. Downward vertical migration triggered under the three stresses

Vertical migration was assessed by tracking $F_0$ measurements under the hydrogen peroxide-, plasma-, and high-light-induced ROS stresses. $F_0$ serves as a proxy for biomass, effectively reflecting the motile behavior of benthic diatoms **[17]**. Indeed, in sediment-free biofilm samples subjected to the hydrogen peroxide, plasma, and high-light treatments, $F_0$ levels did not decrease compared to dark-adapted control replicates (**Supplementary Table 3 and Table 4**). In addition, $F_v/F_m$ values remained stable for the hydrogen peroxide and plasma treatments. Altogether, these results rule out potential Chlorophyll a bleaching and the induction of karyostrophy under any of the three stresses **[32,33]**.

Regardless of how ROS were produced or applied, the *P. strigosum* dominated biofilms sensed both metabolic and environmental oxidative stresses, responding with a marked downward migration (**Fig. 2**). This was quickly followed by a gradual recovery of surface biomass under all stresses as soon as the stress ended. The intensity of the behavioral response varied according to the type of stress, with exogenous ROS triggering the strongest effect. In particular, the hydrogen peroxide treatment induced a clear and rapid on/off switch-migration within ~1 min (experimental observation), with a 75.5% decrease in F0, followed by the plasma exposure causing a 35.6% decrease and an 8.9% reduction under the high-light stress (**Table 2**). Within this framework, the fastest return of biomass at the sediment surface occurred after the hydrogen peroxide exposure, with a +116% F0 min$^{-1}$ recovery slope, compared with +67% under high-light and +47% after the plasma stress, respectively.

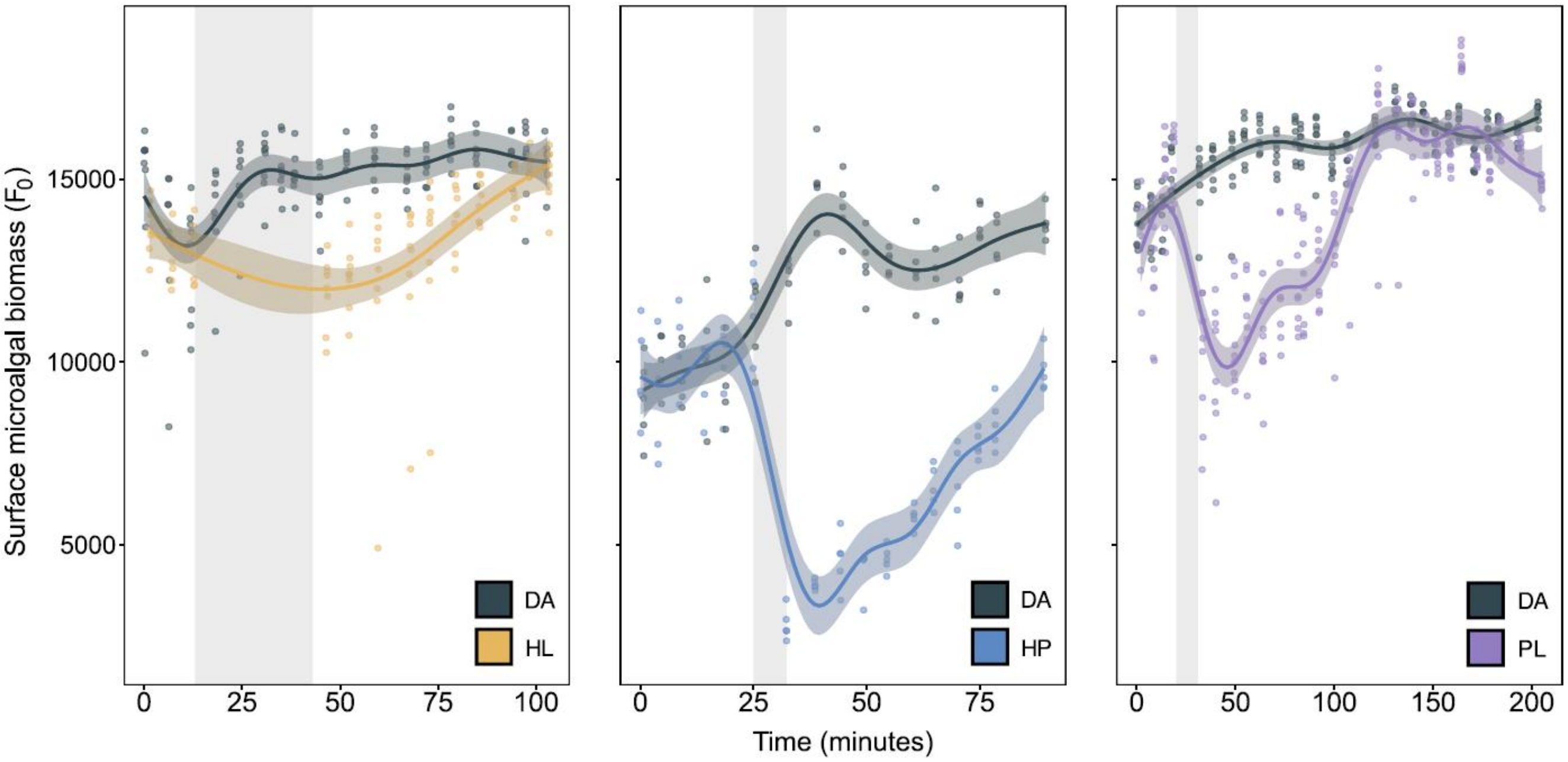

*Figure 2. Downward migration of microphytobenthos under the three oxidative stresses. Temporal monitoring of surface microalgal biomass ($F_0$, arbitrary unit) for sediment-associated biofilms under the three oxidative stresses: High-light (HL; n = 8), hydrogen peroxide (HP; n = 5), and cold atmospheric plasma (PL; n = 8), compared to dark-adapted control biofilms (DA). Shaded grey areas indicate the stress period. All measurements were taken in the morning, during the upward migration phase of microphytobenthos.*

*Table 2. Extent of the behavioral response of the microphytobenthic surface biomass. Table reporting changes in F0 (data from Fig. 2), at the sediment surface (average ± SD, arbitrary unit), before and after treatments, for the sediment-associated biofilm samples (n=8 for high-light and cold atmospheric plasma; n=5 for hydrogen peroxide). Dark-adapted control (1), (2) and (3) correspond to samples measured in parallel to the high-light, hydrogen peroxide, and cold atmospheric plasma experimental treatments, respectively. The Δ After-Before (a.u.) and Changes (%) columns summarize the temporal evolution of surface $F_0$. The post-stress recovery slope is calculated using a linear regression model applied to F0 fluorescence kinetics over 60 min after stress cessation, as an indicator of microphytobenthos resilience. The slope reflects the rate of F0 variation per minute during recovery ($F_0$ min$^{-1}$).*

| Treatments | $F_0$ Before (a.u.) | $F_0$ After (a.u.) | Δ After-Before (a.u.) | Changes (%) | Recovery slope ($F_0$ min$^{-1}$) |
|---|---|---|---|---|---|
| Dark-adapted control(1) | 12 752 (1606) | 14 540 (787) | +1788 | +14% | +11 |
| High-light | 13 080 (807) | 11 916 (993) | −1164 | −9% | +67 |
| Dark-adapted control(2) | 11 393 (1457) | 12 232 (703) | +839 | +7% | −2 |
| Hydrogen peroxide | 11 546 (1100) | 2824 (433) | −8722 | −76% | +116 |
| Dark-adapted control(3) | 15 263 (726) | 14 906 (945) | −356 | −2% | +17 |
| Cold atmospheric plasma | 15 496 (843) | 9979 (1801) | −5517 | −36% | +47 |

## 3.3. Divergent photophysiology: exogenous ROS oxidize pigments with limited photo-synthetic impairment vs. high-light-driven photoinhibition

To complement the finding that microphytobenthic vertical migration behavior may be triggered by ROS, we assessed the metabolic cost associated with these stresses by examining the functional response of the photosynthetic apparatus. To this end, key photosynthetic parameters and lipophilic pigment data were monitored in both communities. Unlike sediment-associated biofilms, the sediment-free biofilms capture the impact on the most complete and intact community possible, as supported by the sequencing results, without the possibility of mitigation through migratory responses.

Before stress exposure, all replicates across all treatments showed high initial $F_v/F_m$ values (>0.67) and high initial α-slope values (>0.32), indicating that the cells were in an optimal physiological state. In both communities, post-exposure photosynthetic efficiency values remained high under all three stresses. This was particularly true for exogenous stresses in sediment-free biofilms, where despite the inability to mitigate stress through migration, the cells maintained high $F_v/F_m$ and α-slope values (>0.66 and >0.33, respectively). Taken together, these results support that, as for the high-light treatment, the intensities of the hydrogen peroxide and plasma stresses remained within the photophysiological tolerance range of the cells despite their inability to migrate, thereby enabling reliable interpretation of the data.

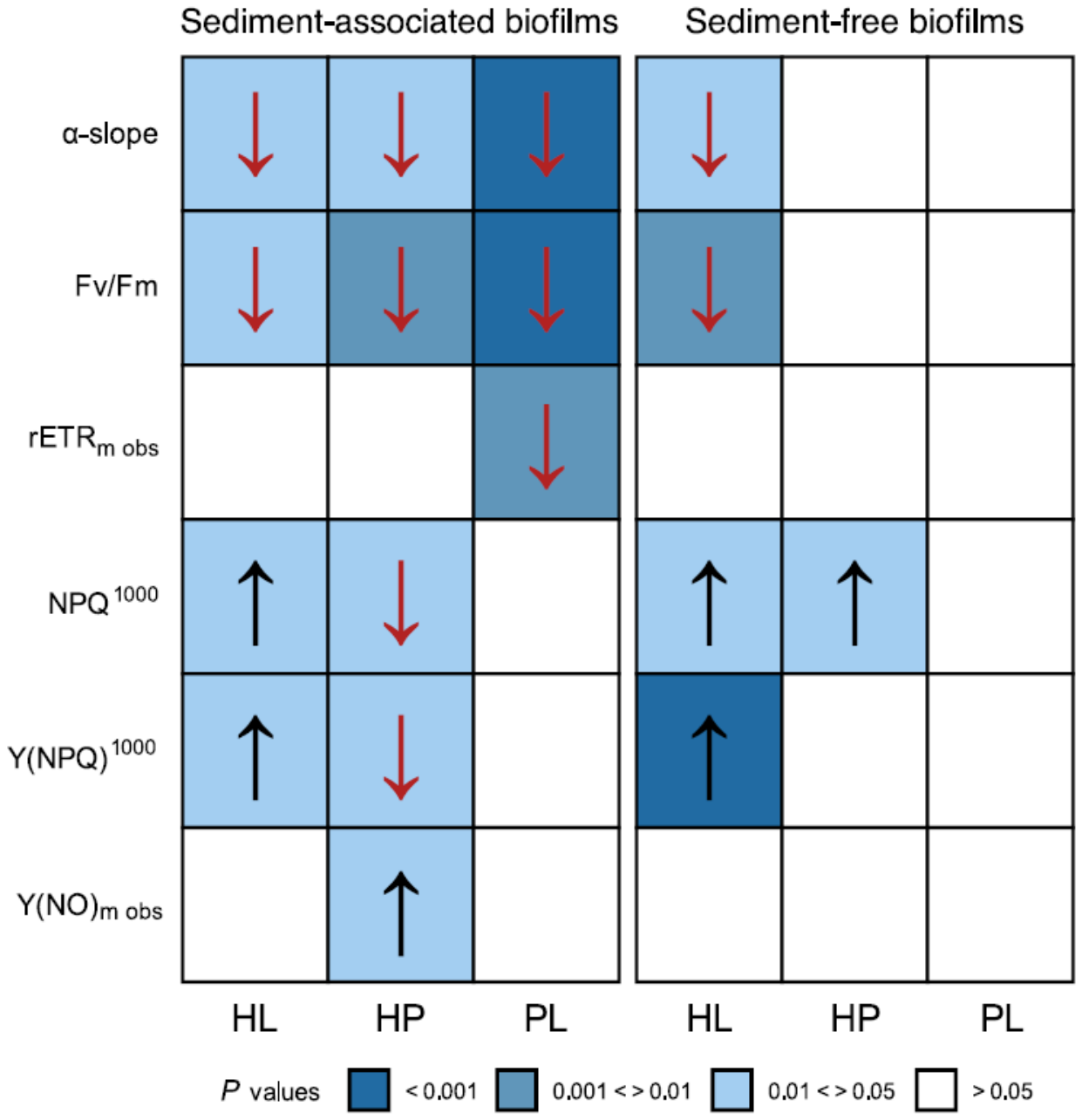

***Figure 3. Heatmap of net photophysiological responses under oxidative stresses. Net impact of high-light (HL), hydrogen peroxide (HP), and cold atmospheric plasma (PL) treatments on photosynthetic parameters. Welch's two-sample t-test (n=3) was used to calculate the stress-induced changes $\Delta_{After\text{-}Before}$, relative to dark-adapted control samples ($\Delta_{After\text{-}BeforeDA}$), in order to correct for the natural photophysiological variation observed under the darkness treatment. Arrows indicate the direction of the change induced by each stressor compared to dark-adapted control biofilms. α-slope (arbitrary unit). $F_v/F_m$ (unitless) corresponds to the maximum quantum yield of photosystem II photochemistry under dark-adapted state. $rETR_{mobs}$ (arbitrary unit) and $Y(NO)_{mobs}$ (unitless) refer to the maximum values observed within the measured light curve range (without extrapolation). NPQ1000 (unitless) and Y(NPQ)1000 (unitless) correspond to the maximum values measured after the final 1-min light curve step at 1000 µmol.photons.m$^{-2}$.s$^{-1}$ (without extrapolation). All fluorescence measurements were performed on dark-adapted samples for both sediment-associated (left panel) and sediment-free biofilm communities (right panel). See Supplementary Table 3 and 4 for detailed values.***

The sediment-associated biofilm community showed a general decrease in photosynthetic efficiency under all three stresses, characterized by a significant drop in the α-slope and $F_v/F_m$ values (Welch's two-sample t-test, $P < .05$). An effective photoprotective response was observed exclusively after the high-light exposure, with increases in $NPQ^{1000}$ and $Y(NPQ)^{1000}$ values, indicating controlled dissipation of excess energy (**Fig. 3**; **Supplementary Fig. 3**). Conversely, stress exposure in sediment-free biofilms revealed a marked photophysiological impact only under the high-light treatment, with a drop in $F_v/F_m$ values, along with a strong activation of the effective photoprotection Y(NPQ) and an increase in unregulated photoinhibition $Y(NO)_{mobs}$. The increase in this uncontrolled dissipation component reflects an excess of irradiance absorbed by the cells under the high-light treatment, estimated at ~519 µmol.photons.m$^{-2}$.s$^{-1}$ by Qphar and recalculated at ~448 µmol.photons.m$^{-2}$.s$^{-1}$ when accounting for the cell pigment packaging effect. The hydrogen peroxide and plasma exposures did not significantly alter the state of the photosynthetic electron transport chain, as particularly evident when comparing the photosystem II quantum yield response curves (**Fig. 4**). Nevertheless, the exogenous plasma stress appeared more impactful than the hydrogen peroxide, highlighted by a decrease in $rETR_{mobs}$ in both sediment-associated and sediment-free biofilm communities.

Microphytobenthic lipophilic pigment analyses revealed diatom-typical profiles in all samples, with dark-adapted sediment-free biofilms providing the most representative unstressed baseline composition: Chlorophyll *a* and its derivatives (58.3%), Fucoxanthin and its derivatives (28%), Chlorophyll *c2* (6.7%), Diadinoxanthin + Diatoxanthin (2.4%), ββ + βε-carotene (1.3%) (**Supplementary Fig. 4**). Both communities were characterized by a clear difference in their pigment compositions (ANOSIM, R = 1; *P* = 1e-04, **Supplementary Fig. 5A**). This was due to higher relative proportions of Chlorophyll *a*, Fucoxanthin, and Chlorophyll *c2* in sediment-free biofilms, whereas sediment-associated biofilm samples exhibited an increase in degraded forms such as Pheopigments, Unknown-carotenoids, or Fucoxanthin-like. Within both communities, stresses significantly affected pigment profiles (sediment-associated biofilms ANOSIM R = 0.82, P = 1e-04, **Supplementary Fig. 5B** and sediment-free biofilms ANOSIM R = 0.73, P = 1e-04, **Supplementary Fig. 5C**), which explain 84% and 67% of the total inertia observed in the sediment-associated and sediment-free biofilm samples, respectively (**Supplementary Fig. 6**). SIMPER analyses revealed a stronger impact of exogenous stresses on pigment composition compared to the high-light stress in both communities, with a greater number of significantly affected pigments relative to dark-adapted control samples (SIMPER, P<.05). This was particularly evident in sediment-associated biofilm samples, where 3, 17, and 39 pigments were significantly affected by the high-light, hydrogen peroxide, and plasma treatments, respectively, compared to 10, 15, and 16 pigments for sediment-free biofilms. This differential impact of the stressors was mainly driven by Chlorophyll a-derivatives rather than Pheopigments (**Fig. 5B** and **Fig. 5C**), suggesting that the hydrogen peroxide and plasma exposures caused a broader and untargeted effect, with direct oxidation of numerous pigments despite short exposure times, whereas the high-light treatment represented a more targeted stress. The latter was particularly characterized by the activation of the xanthophyll cycle, with significant increases in the de-epoxidation state, especially in sediment-free biofilms, where the average ratio exceeded 46% (**Fig. 5A**, concentrations detailed in **Supplementary Table 5**). Finally, sediment-free biofilms showed the same impact dynamics on Chlorophyll a-derivatives as sediment-associated biofilms, confirming that the stresses differentially affected the surface motile diatoms themselves.

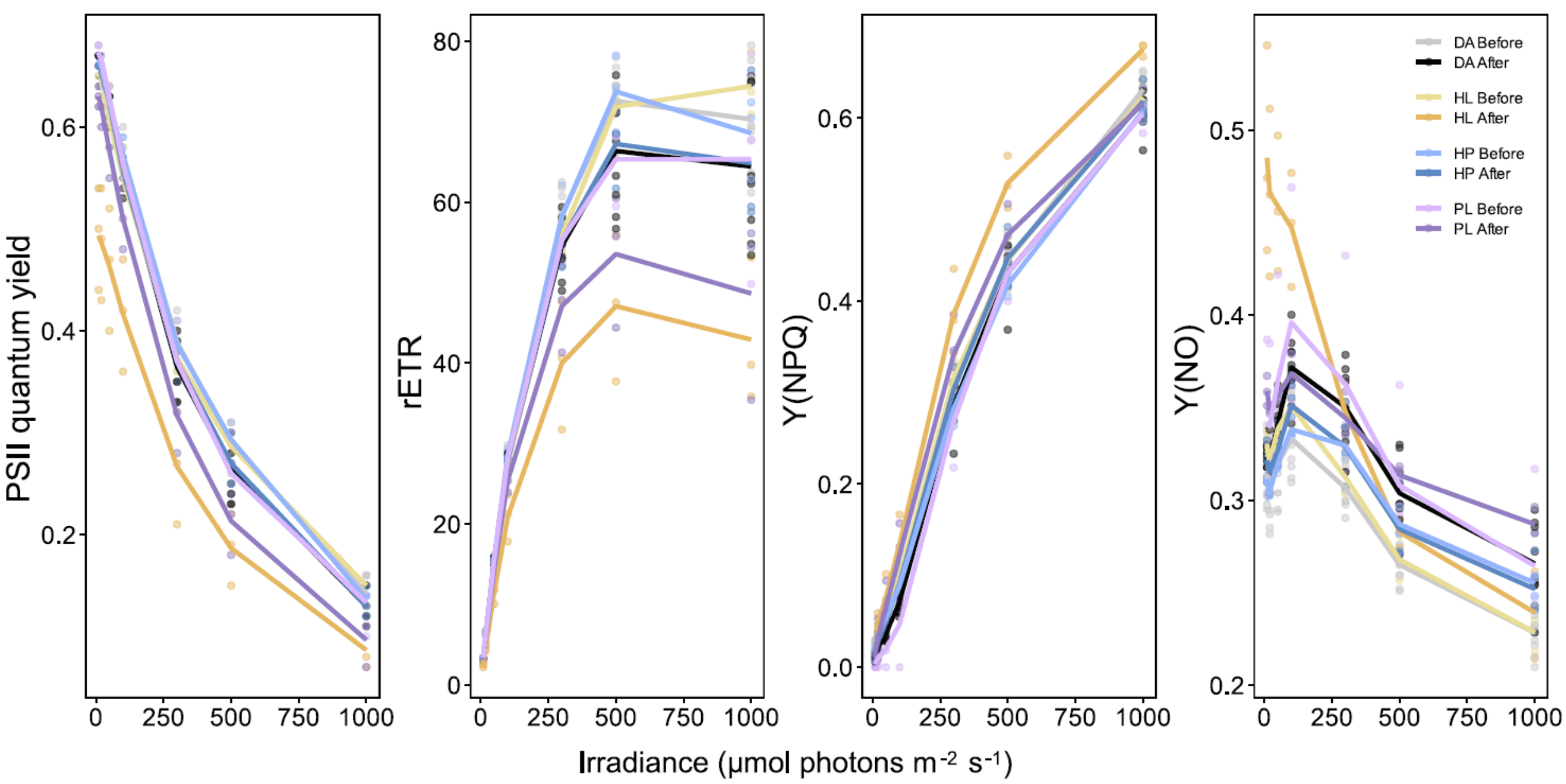

***Figure 4. Photosynthetic efficiency profiles under treatments. Photosystem II (PSII) quantum yield (unitless), rETR (arbitrary unit), Y(NPQ) (unitless), and Y(NO) (unitless) response curves of sediment-free biofilm community, under the three stresses: High-light (HL), hydrogen peroxide (HP), and cold atmospheric plasma (PL), compared to dark-adapted control biofilms (DA). All photosynthetic parameters measured with n = 3.***

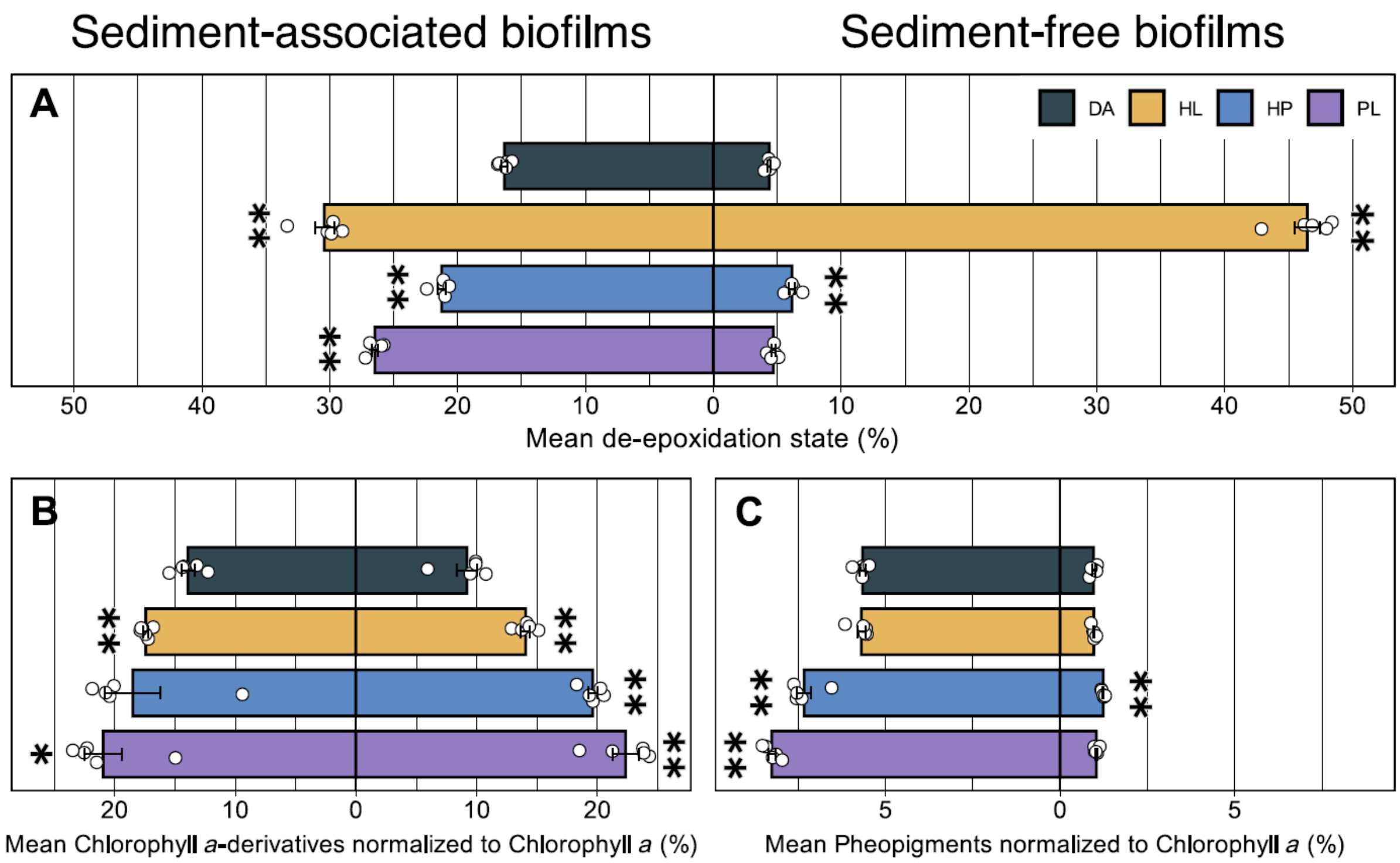

***Figure 5 Impact of treatments on pigment content. Barplots showing variations in pigment ratios (%) under the following treatments: Dark-adapted control (DA), high-light (HL), hydrogen peroxide (HP), and cold atmospheric plasma (PL), for both sediment-associated (left panel) and sediment-free biofilm communities (right panel); A. Barplots of the de-epoxidation state (%); B. Barplots of chlorophyll a-derivatives pigment ratios; C. Barplots of Pheopigments pigment ratios. Statistical differences were assessed by the Mann–Whitney U test between the dark-adapted control biofilm samples and each stress (n = 5; P < .05 *, < .01 **).***

# 4. Discussion

## 4.1. High-light-induced vertical migration in microphytobenthos

Pigment profiles provided a chemotaxonomic signal typical of diatoms, characterized by high fucoxanthin content together with Chlorophyll c2 and Chlorophyll a [28], which corroborated the sequencing results (Fig. 1E). Both communities were characterized by a biovolume dominated by benthic diatoms, especially *P. strigosum*, along with other autotrophs and meiofauna (e.g. nematodes, copepods, and ostracods), corroborating previous microscopic observations [49] and literature on microphytobenthos in similar environments [50,51]. In line with these observations, the motile behavioral response, pigmentary mobilization, and photophysiological signals can be largely attributed to the benthic diatoms dominating the surface biovolume, with *P. strigosum* playing a predominant role.

In sediment-associated biofilm experiments, exposure of the microphytobenthos to 1500 µmol.photons.m$^{-2}$.s$^{-1}$ for 30 min induced a moderate decrease in $F_0$, indicating a shift in biofilm dynamics from upward to downward migration (Fig. 2). This response was reversible, with biomass returning to the surface after the treatment ended. Vertical migration in benthic diatoms is governed by self-sustained circadian and tidal rhythms that require regular resetting to remain synchronized [20,52]. These rhythms are further modulated by environmental cues, allowing rapid and temporally adjustable responses [18,21,23]. The induction of a high-light-driven vertical migration strategy through muddy substrata was expected and is well documented in many autotrophic microorganisms, including *P. strigosum*, highlighting a relatively common evolutionary advantage [17,20,26]. To maintain optimal cellular performance and cope with ROS, this migratory response operates in synergy with other photoprotective mechanisms that collectively sustain cellular fitness [6].

## 4.2. Migration and photoprotection act synergistically under high-light-induced ROS

In parallel to the downward migration, a decline in photosynthetic efficiency was observed in the high-light-treated sediment-associated biofilm samples, suggesting that vertical migration did not maintain short-term photosynthetic performance. This decline was likely caused by an increase in Y(NPQ) due to targeted mobilization of the Diadinoxanthin-Diatoxanthin cycle (Fig. 3 and Fig. 5A), combined with slight photoinhibition, consistent with previous reports [26,53]. Irradiance is known to generate both endogenous and exogenous ROS-high-oxidative-potential molecules that can cross membranes and act as signaling agents across multiple subcellular compartments [6,9,54–59]. Diatoms must cope with them, as excessive accumulation poses deleterious risks, including the oxidation of essential photosynthetic components [10, 11]. To prevent ROS formation, high-light can rapidly trigger effective Y(NPQ) supported by the Diadinoxanthin + Diatoxanthin xanthophyll pool, where overstimulation of the photosynthetic electron transport chain induces lumen acidification, upregulating the pH-sensitive diadinoxanthin de-epoxidase enzyme [15, 30,60]. Here, both vertical migration and xanthophyll cycle activation likely limit ROS production and accumulation, indicating that all components of a typical and effective photoprotection strategy are engaged, as expected for benthic intertidal diatoms under photooxidative stress. In addition, diatoms actively regulate cellular redox balance under light-induced ROS via both enzymatic and non-enzymatic antioxidant capacities, including upregulation of superoxide dismutase and ascorbate peroxidase [37,39,61], mobilization of antioxidant non-protein thiols and glutathione pools [62], or by direct ROS scavenging by carotenoids [29,31,38].

Although Chlorophyll a fluorescence is generally a reliable indicator of cellular physiological state under various abiotic stresses [63], the presence of a sedimentary environment in sediment-associated biofilm samples can alter the signal (e.g. cell reorientation, effective light exposure, or downward migration, potentially leaving fewer viable cells at the surface). Consequently, sediment-associated biofilms photosynthetic parameters must be interpreted with caution [45,64]. The sediment-free biofilms results help to overcome this potential bias, thereby clarifying the synergy between the need for migration, the xanthophyll cycle activation, and the photosynthetic efficiency. Cells from sediment-free biofilms, unable to adjust their vertical position, displayed a stronger increase in the de-epoxidation state, as a high-light-compensatory response (Fig. 5A). Enhanced xanthophyll cycle mobilization is a well-described compensatory response in diatoms with impaired motility, as shown under light exposure combined with latrunculin motility inhibitors [14,26,65]. In natural environments, this trade-off becomes evident when examining the de-epoxidation state of epipsammic diatoms, which inhabit brighter environments where vertical migration is limited, resulting in a stronger response than in epipelic diatoms [65-67]. Despite the xanthophyll cycle response in sediment-free biofilms, an increase in Y(NO) indicates that the stress still induced a slight unregulated photoinhibition from the lowest irradiances of the light curves (Fig. 4).

The high-light treatment induced finely tuned and metabolically controlled cellular adjustments, involving subtle pigment remodeling as part of a photoprotective strategy, which works in synergy with vertical migration behavior. However, although this targeted response is likely driven by intracellular variations (e.g. metabolic imbalances resulting from ROS accumulation primarily due to light, and potentially in combination with other stresses, such as temperature, pH, exposure to toxic substances, or nutrient availability), migration may also result from environmental oxidative signals [58]. These include exogenous high-light-ROS generated by photooxidation of dissolved organic matter in muddy substrata [56,58,68], and ROS actively or passively released by neighboring cells in the biofilm, leading to their local accumulation [13,54,69]. Moreover, precipitation can deliver water containing up to 85 µM $H_2O_2$ [58]. All these conditions in intertidal environments can rapidly and transiently increase exogenous oxidative stress for microphytobenthos, an environmental scenario not far from the hydrogen peroxide and plasma treatments, although these stresses involve higher concentrations than those typically measured in tidal zones [56].

## 4.3. Exogenous ROS induce untargeted oxidative stress on diatoms

As for high-light-induced photooxidation, exogenous ROS generated by the hydrogen peroxide and plasma stresses, including $H_2O_2$, likely interacted with sedimentary dissolved organic matter, triggering chemical reactions. These reactions could alter oxidative conditions in sediment samples, including potential acidification that may explain the increase in the de-epoxidation state. This xanthophyll cycle activation in sediment-associated biofilms appeared to occur without development of regulated Y(NPQ) (**Supplementary Fig. 3**). As previously discussed, downward vertical migration into the sediment can strongly interfere with fluorescence measurements, including underestimation of NPQ [45,64]. In addition, migration of viable cells into deeper sediment layers may leave senescent or physiologically impaired cells at the surface, further altering the residual fluorescence signal. This interpretation is supported by the higher relative abundance of Pheopigments observed in sediment-associated biofilms compared to sediment-free biofilm samples (**Fig. 5C**), along with the increase in Y(NO), concomitant with the apparent loss of Y(NPQ) development capacity. Due to the substantial downward migration of biomass, the reliability of surface fluorescence measurements in sediment-associated biofilm samples is intrinsically compromised, and it is therefore not possible to definitively conclude whether the observed signal reflects a reduced and physiologically altered surface population or whether the induced diatoxanthin is preferentially used in antioxidant functions rather than photoprotective NPQ [38].

Exogenous stresses in both communities exhibited a stronger untargeted oxidizing capacity, leading to extensive and differential oxidation of the pigment system. The presence of sediment, potentially creating altered oxidative conditions, combined with an associated pool of degraded Pheopigments, could explain the higher Between-Class Analysis explained inertia values observed for sediment-associated biofilms (**Supplementary Fig. 6**). The global impact of the hydrogen peroxide and plasma treatments is particularly reflected in the oxidation of Chlorophyll a into Chlorophyll a-derivatives (**Fig. 5B**), suggesting a random and direct oxidation of lateral chains of Chlorophyll a molecules [70]. Some of these alternative forms of Chlorophyll a may have cellular functions [9,29,71], and their accumulation may serve as an indicator of environmental conditions [5]. Despite the plasma treatment inducing a higher accumulation of Chlorophyll a-derivatives, indicating more detrimental conditions than the hydrogen peroxide exposure, both exogenous stresses resulted in high post-stress photosynthetic performance. This supports the relevance of the applied stress intensities, as structural modifications in part of the Chlorophyll a pool did not prevent the maintenance of light-harvesting functions, even when migration was impaired.

## 4.4. Vertical migratory behavior triggered by exogenous ROS

Our results showed that brief exposure to hydrogen peroxide and plasma-ROS had a moderate impact on photophysiology (Fig. 4), while simultaneously triggering a rapid downward vertical migration (Fig. 2). Reported migration velocities are highly variable, typically ranging from 1 to 30 µm $s^{-1}$ for horizontal movements, and generally lower for vertical displacements, generally between 0.17 and 0.28 µm $s^{-1}$ [17,20,21,24]. Based on these values, diatoms could theoretically migrate vertically by ∼102 µm and 168 µm under our 10-min stresses, reaching muddy layers where light availability is reduced by ∼66% [65], which is consistent with the strong decrease observed in F0. In both cases, stresses did not compromise the motility of diatoms, which rapidly initiated an upward migration once the stresses ended, as shown by post-stress recovery slope (**Table 2**). This plasticity in behavioral response, which adjusts in real time to changing surface conditions, was nevertheless more affected under the plasma exposure, with a slower return of biomass, suggesting a harsher stress. ROS, such as $H_2O_2$, are described in literature as secondary messengers that mediate signal transduction and activate defense mechanisms during adaptation to both biotic and abiotic stress [5,6,8,59], including motile capacities in microalgae [25,55,57,72,73]. The hypothesis that ROS are one of the components inducing migration under a composite stress such as cold atmospheric plasma is corroborated by the strong migration observed under the hydrogen peroxide positive control, which specifically isolates the effect of $H_2O_2$. This molecule has been widely used in previous studies to investigate its role as a driver in various cellular pathways in microalgae [59,72,74–77], as exogenously applied $H_2O_2$ can accumulate intracellularly via transmembrane diffusion through aquaporins [78]. The results presented here demonstrate its key role as a strong on/off trigger of the transient motile response. Moreover, the behavioral response did not reflect the impact of exogenous stresses on the pigment system, as the strongest mobilization occurred under the hydrogen peroxide treatment rather than under plasma.

Together, these results suggest that vertical migration could be triggered by ROS, especially $H_2O_2$, without necessarily being associated with a decline in photosynthetic performance. In other words, the presence of ROS, even in darkness and without impaired photosynthetic capacity, is sufficient to trigger downward motility. This triggering of migration occurs without inducing an effective controlled photoacclimatory Y(NPQ) response involving the xanthophyll cycle. Exogenous ROS likely acted through a sophisticated signal transduction network to trigger this rapid and essential motility response of the microphytobenthos under harmful oxidative environmental conditions [6,8]. The precise cellular pathways by which $H_2O_2$ act, either directly or indirectly, to trigger this behavior remain uncertain in benthic diatoms and require further investigation. Nevertheless, certain studies in the literature provide a solid foundation for preliminary research.

## 4.5. Do ROS induce calcium signaling underlying motility?

In benthic pennate diatoms, gliding motility is driven by an actin-based cytoskeletal system organized along the apical axis. This system transmits force through transmembrane complexes to mucopolysaccharide microfibrils secreted via the raphe slit, thereby enabling adhesion and traction on substrata [20,22]. ROS can plausibly act on several components of the gliding system, including cytoskeletal reorganization, adhesive mucilaginous compounds, ATP availability, gene expression, or by generating oxidized metabolites that function as infochemicals and secondary messengers in specific signaling pathways (e.g. oxylipins, thiol compounds) [12,22,79-83]. Nevertheless, a migration behavior primarily controlled by ROS-calcium signaling remains one of the most prominent hypotheses, both of them being central components of microalgal adaptation to environmental stressors [84].

Recent studies have shown that disruption of the transt hylakoid proton gradient inhibits motility in benthic diatoms, potentially through effects on calcium homeostasis, thereby suggesting a broader link between behavior and high light-driven photoprotection [77,85]. Our results provide experimental evidence that diatom migration can be triggered in darkness and independently of an effective Y(NPQ) response, further refining the regulatory cascade underlying this behavior by linking ROS to an "oxidotactic" motility that may involve calcium signaling. Elevations of stromal $Ca^{2+}$ have been reported in response to intracellular ROS accumulation under high light and after exogenous $H_2O_2$ exposure [76,77]. In plants, both endogenous $H_2O_2$ accumulation and exogenous $H_2O_2$ exposure induce $Ca^{2+}$ influx via calcium channel activation [86], an influx implicated in controlling diatom migration [87,88] and, more recently, contributing to gliding motility through EukCatAs ion channels that regulate cytosolic $Ca^{2+}$ levels [89]. In addition, it has been shown that certain membrane receptor kinases can directly sense extracellular $H_2O_2$ and initiate $Ca^{2+}$ channel opening without a prior intracellular $H_2O_2$ accumulation [90]. In the environment, exposure to copper promotes oxidative stress and intracellular $Ca^{2+}$ perturbations alongside active gliding motility [25], and new evidence has identified a bistability in phytoplanktonic microalgae, in which vertical migration switches are driven by a ROS-induced oxidative stress threshold [72].

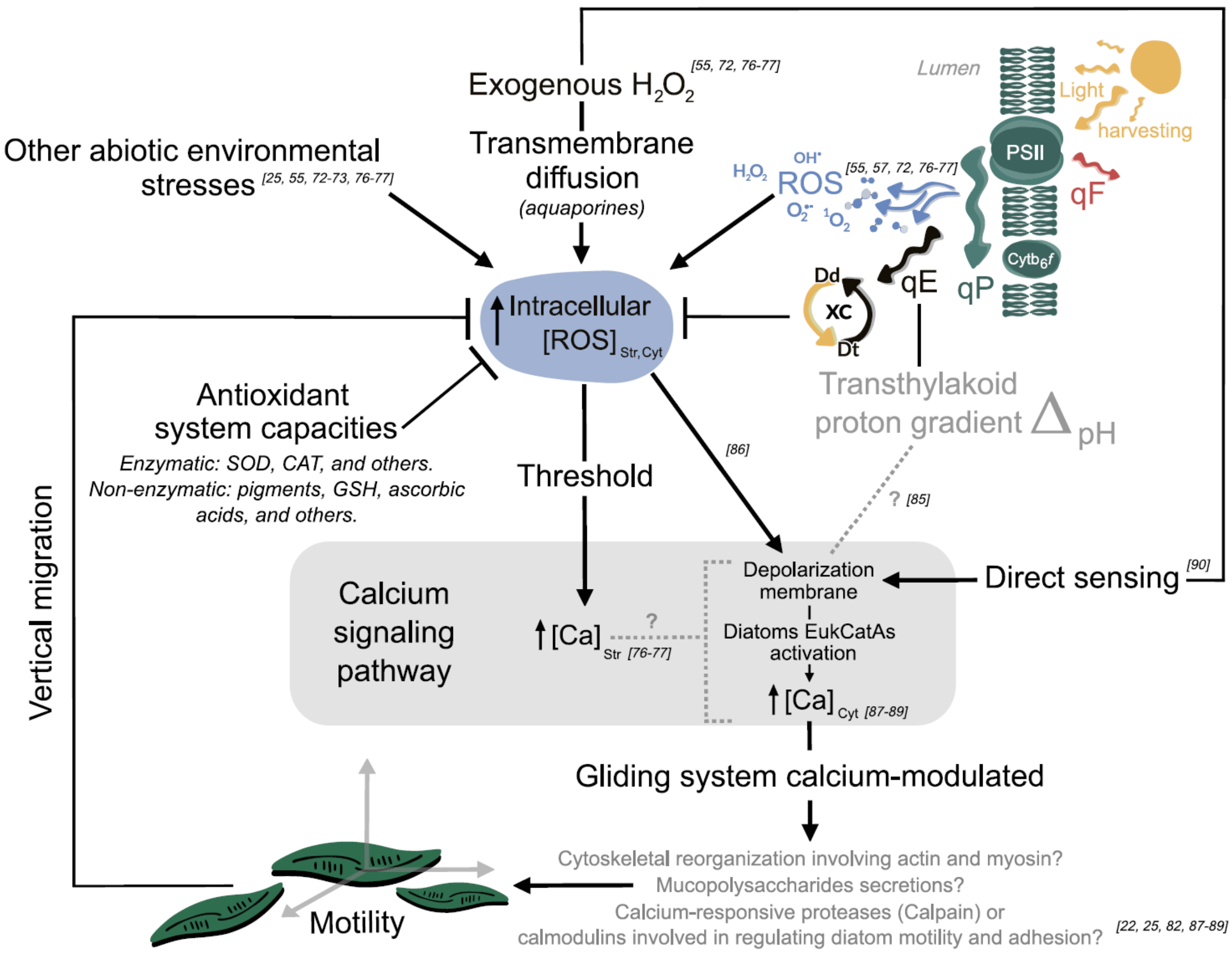

***Figure 6** Putative integration of ROS-calcium cellular signaling underlying stress-induced downward vertical migration. Summary of the ROS-calcium-motility interplays illustrating a hypothetical signaling model linking the observed motile behavior to the applied stresses. Black arrows and their direction indicate relationships supported by the literature between specific components. Grey dashed lines represent hypothetical interactions. Lines (⊥) indicate processes involved in regulating intracellular ROS concentrations. Abbreviations used in the figure: "qF" fluorescence quenching; "qP" photochemical quenching; "qE" energy-dependent quenching; "str" stromal; "cyt" cytosolic. Numbers in brackets indicate references.*

Our study does not provide a definitive conclusion on the full role of ROS, such as $H_2O_2$. Nevertheless, these results, supported by recent literature, help to identify the major putative components underlying the cellular signaling of this behavior (**Fig. 6**). Although detailed information on the ROS-Calcium-Motility interplays is still lacking, particularly regarding how the calcium acts on the locomotion machinery pathways, we propose that certain rapid responses to environmental cues, such as a vertical migratory behavior switch observed in this experiment, are mediated, at least in part, by a ROS-sensing system.

# 5. Conclusion

All results provide strong evidence for the key role of ROS, particularly $H_2O_2$, as a critical signaling molecule triggering the rapid motile behavior in diatom-dominated microphytobenthic biofilms. The strongest downward migration occurred under exogenous stresses that minimally impacted photosynthetic efficiency, despite a broad and untargeted oxidative stress on pigments. Moreover, these exogenous stresses were unable to induce an effective, controlled photoprotective response, as reflected by Y(NPQ). This demonstrates that exposure to $H_2O_2$ alone, in the absence of any light cue, is sufficient to trigger a plastic motility. Such a response suggests a signaling pathway that may operate independently from those governing photophysiology, thereby broadening the ecological relevance of vertical migration in a multi-stress environmental context. This stress-mitigation strategy can act synergistically with pigments under high irradiance in natural environments, as supported by the enhanced activation of the xanthophyll cycle when migration was restricted. Together, the findings corroborate previous studies showing that vertical migration is an important adaptive trait allowing diatoms to regulate homeostasis and support long-term physiological performance.

By demonstrating the importance of ROS sensing in shaping microphytobenthic motility, these results indicate that benthic diatoms are able to monitor both metabolic and environmental ROS. This ROS-mediated early warning sensing system likely represents an adaptive strategy that enables efficient surveillance of environmental conditions, allowing cells to rapidly prevent deleterious conditions, maintain cellular fitness, and ultimately support the ecological success of diatoms in highly variable environments. This offers new insights into how oxidative stress regulates autotrophic life and more broadly highlights its key role within the aquatic microbiome. The response to oxidative challenges likely relies on a complex signaling cascade that requires further in-depth investigation to identify potential effectors in ROS-motility signaling and clarify how triggering this behavior allows cells to cope with oxidative stress in natural environments.

# 6. Acknowledgements

Our thanks go to Johann Lavaud for his valuable discussions on photophysiology, and Cyril Noël for his assistance with the SAMBA workflow, which enabled the analysis of the metabarcoding data. We are also thankful to Manon Leroux for her essential assistance in the laboratory during the experiment. We acknowledge the chromatography and mass spectrometry facility at the Concarneau Marine Station (PtSMB-MNHN) for providing access to the HPLC.

# 7. Author contributions

AD (Conceptualization, Methodology, Investigation, Data curation, Formal analysis, Visualization, Writing—original draft), BJ (Data curation, Formal analysis, Validation, Writing—review & editing), TR (Data curation, Methodology, Formal analysis), TD (Resources, Methodology, Writing—review & editing), and CH (Supervision, Methodology, Data curation, Formal analysis, Validation, Writing—review & editing).

# 8. Supplementary materials

Supplementary material is available at The ISME Journal online.

# 9. Conflicts of interest

The authors declare no conflict of interest.

# 10. Funding

This work was funded by the Institut de l'Océan from the Sorbonne University alliance and the Muséum national d'Histoire naturelle.

# 11. Data availability

All scripts used to generate the figures and perform the analyses are available on GitHub (https://github.com/adesparmet/ROS-Triggered-Diatom-Biofilm-Migration), and are also archived on Zenodo (https://zenodo.org/records/17813206). The datasets generated and analyzed during the current study are available in the Zenodo repository (https://doi.org/10.5281/zenodo.15835853). The raw sequencing data have been deposited in the European Nucleotide Archive (ENA) under project accession number ERA33152396 (Experiments ERX14500286-ERX14500369; Runs ERR15095684-ERR15095767) and can be accessed via the ENA portal (https://www.ebi.ac.uk/ena/browser/view/ERA33152396). The raw chromatogram files used for pigment analyses are available in the Zenodo repository (https://doi.org/10.5281/zenodo.15847025).